\documentclass[12pt]{article}
\usepackage{latexsym}
\usepackage{amsmath,,calrsfs}
\usepackage{amsfonts}
\usepackage{amssymb}
\usepackage{amscd}
\usepackage{fancybox}
\usepackage{cite}
\usepackage{amsmath,amsfonts,amsbsy}
\usepackage{pstricks,pst-node}
\usepackage[small,bf,hang]{caption2}
\usepackage{psfrag}

\usepackage{float}

\psset{unit=1.3cm,linewidth=.5pt,radius=.2}  

\usepackage{enumitem}                       
\usepackage{multirow}                     
\usepackage{float}                          
\usepackage{lscape}                         
\usepackage{bm}

\def\hybrid{\topmargin -20pt    \oddsidemargin 0pt
        \headheight 0pt \headsep 0pt
        \textwidth 6.25in       
        \textheight 9.5in       
        \marginparwidth .875in
        \parskip 5pt plus 1pt   \jot = 1.5ex}

\hybrid

\newcommand{\beq}{\begin{equation}}
\newcommand{\eeq}{\end{equation}}
\newcommand{\bi}{\begin{itemize}}
\newcommand{\ei}{\end{itemize}}
\newcommand{\bt}{\begin{tabular}}
\newcommand{\et}{\end{tabular}}
\newcommand{\bc}{\begin{center}}
\newcommand{\ec}{\end{center}}

\newcommand{\be}{\begin{equation}}
\newcommand{\ee}{\end{equation}}
\newcommand{\bea}{\begin{eqnarray}}
\newcommand{\eea}{\end{eqnarray}}
\newcommand{\ba}{\begin{array}}
\newcommand{\ea}{\end{array}}

\def\bbox{{\,\lower0.9pt\vbox{\hrule \hbox{\vrule height 0.2 cm
\hskip 0.2 cm \vrule height 0.2 cm}\hrule}\,}}
\newcommand{\dsl}{\pa \kern-0.5em /}

\begin{document}

\begin{titlepage}
\begin{center}

\hfill  DAMTP-2013-8

\vskip 1.5cm

{\Large \bf The Hamiltonian Form of Topologically Massive Supergravity}

\vskip 1cm

{\bf Alasdair Routh$^1$}

\vskip 25pt

{\em $^1$ \hskip -.1truecm Department of Applied Mathematics and Theoretical Physics,\\ Centre for Mathematical Sciences, University of Cambridge,\\
Wilberforce Road, Cambridge, CB3 0WA, U.K. \vskip 5pt }

{email: {\tt A.J.Routh@damtp.cam.ac.uk}} \\

\end{center}

\vskip 0.5cm

\begin{center} {\bf ABSTRACT}\\[3ex]\end{center}

We construct a ``Chern-Simons-like'' action for $\mathcal{N} = 1$ Topologically Massive Supergravity from the Chern-Simons actions of $\mathcal{N}=1$ Supergravity and Conformal Supergravity. We convert this action into Hamiltonian form and use this to demonstrate that the theory propagates a single massive $\left(2, \frac{3}{2}\right)$ supermultiplet.

\end{titlepage}

\newpage
%

\section{Introduction}
\setcounter{equation}{0}

Theories of gravity in spacetimes of three dimensions (3D) have been studied extensively over the past few decades, both as tools for understanding gravity in four or more dimensions and for their own intrinsic interest. It is well-known that in 3D, massless spin-2 particles have no local degrees of freedom, and correspondingly 3D General Relativity (GR) is ``trivial''. One thing this suggests is that studying 3D GR and its quantisation could be a helpful tool on the way to understanding the quantisation of 4D GR, see for example \cite{Deser:1984,Witten:1988hc}. It also suggests that in order to find a ``non-trivial'' 3D gravity theory, one should look at models of interacting massive spin-2 particles. Massive gravity is of general interest\cite{Hinterbichler:2011tt} and understanding the situation in 3D could help the development of more complicated 4D models.

In 3D a massive spin-2 particle has one local degree of freedom, and the first theory of such a particle to be discovered was Topologically Massive Gravity (TMG)\cite{Deser:1981wh}. This theory breaks parity, as a parity invariant theory must have two massive spin-2 particles of opposite helicities. Recently, this has been realised by another massive gravity model, New Massive Gravity (NMG)\cite{Bergshoeff:2009hq, Bergshoeff:2009aq}, which propagates two modes of the same mass and opposite helicities and is parity invariant. The combination of the two models is called Generalised Massive Gravity (GMG) which propagates two modes of different masses and opposite helicities, and TMG and NMG can be realised as limits of this more general model.

The above discussion generalises to supergravity. The 3D massless super-multiplet containing a spin-2 particle as its highest spin state also has no local degrees of freedom, and so like its bosonic counterpart, 3D supergravity is ``trivial''. Supersymmetric counterparts of TMG, NMG and GMG have been found\cite{Andringa:2009yc,Bergshoeff:2010mf,Bergshoeff:2010ui}, and in this paper we will be particularly interested in the $\mathcal{N}=1$ supersymmetric extension of TMG, called $\mathcal{N}=1$ Topologically Massive Supergravity (TMSG) first described in \cite{Deser:1983}. 

3D General Relativity\cite{Achucarro:1987vz, Witten:1988hc} and Conformal Gravity\cite{Horne:1988jf} can both be formulated as Chern-Simons theories of the 3D Poincar\'e Group (or A/dS Group if a cosmological constant is included) and Conformal Group respectively. In this form, their actions are integrals of 3-form Lagrangians constructed from exterior products of 1-forms and exterior derivatives. These 1-forms are the dreibein, spin-connection and in the case of Conformal Gravity extra fields corresponding to special conformal transformations and dilatations. TMG was originally formulated by adding together the action of 3D GR and an alternative gauge-fixed second-order action for Conformal Gravity, but it was recently noticed that one could get the same theory by combining the two Chern-Simons actions\cite{Hohm:2012}, putting TMG into ``Chern-Simons-like'' form. By this we mean that it is similar to a Chern-Simons theory in that it is described by an action which is the integral of a 3-form Lagrangian constructed from 1-form fields and exterior derivatives. However we do not require that the action arises from a group structure as in an actual Chern-Simons theory. Such ``Chern-Simons-like'' actions are worth considering for their interesting and useful properties. They are constructed without the use of a metric, they are relatively simple to work with, and importantly they are first-order. This last point relates to the fact that we treat the spin connection as an independent variable, to be determined by the field equations.

Being first-order makes these actions very easy to put into Hamiltonian form. As any term can only have one time-derivative, after a time/space decomposition the actions are automatically in the form ``$P\cdot\dot{X} - \lambda_iC_i$''. The spacelike components of the 1-form fields can be interpreted as canonical variables $X, P$ while the time components of the fields are always non-dynamical and act as Lagrange multipliers $\lambda_i$ imposing various constraints $C_i$. We can use the Hamiltonian analysis described by Dirac in \cite{Dirac:1950pj} to find any additional constraints which must be included. The ``$P\cdot\dot{X}$'' term defines Poisson brackets between the dynamical variables, and using these elementary Poisson brackets, the Poisson brackets of the constraints can be computed. We can then distinguish so-called first-class and second-class constraints and hence determine the number of physical modes the theory propagates.

In \cite{Hohm:2012}, ``Chern-Simons-like'' actions for TMG, NMG and GMG were found and their constraint structures analysed as described above, agreeing with previous results in \cite{Deser:1991qk,Buchbinder:1992pe,Grumiller:2008pr,Carlip:2008qh,Blagojevic:2008bn,Blagojevic:2010ir,Afshar:2011qw}. The aim of this paper is to extend that work to three different $\mathcal{N}=1$ supergravity theories. We will use a component approach, but superspace methods for Conformal Supergravity and TMSG have recently been studied in \cite{Kuzenko:2012}.

We will first review $\mathcal{N}=1$ Supergravity and Conformal Supergravity in their Chern-Simons forms. The Supergravity action is well-known, and while the first-order Chern-Simons version of the Conformal Supergravity action has almost been constructed several times in the literature\cite{VanNieuwenhuizen:1985ff,Uematsu:1986,Foussats:1992}, the explicit expression given here is perhaps novel. We will then use these theories to construct a ``Chern-Simons-like'' version of TMSG and finally transfer each supergravity theory into Hamiltonian form and analyse the constraint structure to determine the number of degrees of freedom each has.

\section{The Supergravity Theories}
\setcounter{equation}{0}

We will begin our analysis by presenting the Chern-Simons forms of the 3D Supergravity and Conformal Supergravity actions from which we construct TMSG, which will allow us to introduce our conventions. We will then discuss TMSG itself. For a more thorough discussion of the corresponding bosonic theories using the same formulation and conventions, see \cite{Hohm:2012}, of which this paper is an extension\footnote{The field named $h^a$ in this reference corresponds to $-2f^a$ here, which is a more natural scaling in the context of Conformal Supergravity.}.

\subsection{Supergravity}

Let us first recall the Einstein-Cartan formulation of 3D gravity. This model uses the dreibein $e^a = e_{\mu}\,^a dx^{\mu}$, a Lorentz-vector valued one-form which gives rise to the metric $g_{\mu\nu} = \eta_{ab} e_{\mu}\,^a e_{\nu}\,^b$, and the spin connection $\omega_{\mu}\,^a = \frac{1}{2}\epsilon^{abc}\omega_{\mu bc}$, another 1-form which in three dimensions can be dualised as shown. We also dualise the Riemann tensor $R^a = \frac{1}{2}\epsilon^{abc}R_{bc} = d\omega^a + \frac{1}{2}\epsilon^{abc}\omega_b\omega_c$, and define the covariant derivative of a Lorentz-vector valued one-form, $Dh^a = dh^a + \epsilon^{abc}\omega_bh_c$. We implicitly take the wedge product of adjacent forms.

In this notation the standard Einstein-Hilbert action becomes the integral of the Lagrangian three-form

\be
L_{EC} = \frac{1}{2}\sqrt{-g}\,(R - 2\Lambda) \, dx^0dx^1dx^2 = -e_aR^a + \frac{\Lambda}{6}\epsilon^{abc}e_ae_be_c \,.
\ee

This is a ``first-order'' formulation, we consider the spin connection to be an independent variable rather than a fixed function of the dreibein. Varying both of the forms, we get the following Euler-Lagrange equations corresponding to $(e, \omega)$ respectively

\be
R^a -\frac{\Lambda}{2} \epsilon^{abc}e_be_c = 0 \,, \qquad D e^a = 0 \, .
\ee

$D e^a$ is the torsion which the second equation sets to be zero. This determines the spin connection to be the usual one, which we will call $\omega(e)$. The first equation is then equivalent to the Einstein equation, $G_{\mu\nu} + \Lambda g_{\mu\nu} = 0$. Note that this form of the action for GR is first order in time derivatives, which will be important when we look at the Hamiltonian form later.

We can extend this action to a supergravity action by introducing a gravitino, an anti-commuting Majorana spinor valued 1-form, $\psi_{\mu}\,^{\alpha}$. Define the covariant derivative of a spinor valued one-form, $D\chi = d\chi + \frac{1}{2}\omega_a \gamma^a \chi$ where the $\gamma^a$ satisfy $\left\{ \gamma^a, \gamma^b \right\} = 2\eta^{ab}$ with the $(-++)$ convention, and whenever a representation of the $\gamma^a$ is called for we will use a real representation defined in terms of the Pauli matrices

\be
\gamma^0 = i\sigma_2\,, \qquad \gamma^1 = \sigma_1\,, \qquad \gamma^2 = \sigma_3 \,.
\ee

Our supergravity action is

\be
L_{SEC} = -e_aR^a - 2\bar{\psi} D \psi - \frac{\lambda^2}{6} \epsilon^{abc}e_ae_be_c + \lambda e_a \bar{\psi}\gamma^a\psi  \,.
\ee

The cosmological constant is $\Lambda = -\lambda^2$, non-positive in supergravity. Again vary each form to get the Euler-Lagrange equations corresponding to $(e, \psi, \omega)$

\be
R^a + \frac{\lambda^2}{2}\epsilon^{abc}e_be_c - \lambda \bar{\psi}\gamma^a\psi = 0 \,, \qquad D\psi + \frac{\lambda}{2}e_a\gamma^a\psi = 0 \,, \qquad  D e^a - \bar{\psi}\gamma^a\psi = 0 \, .
\ee 

The third equation is the usual supergravity torsion condition, which defines the spin connection in terms of the dreibein and the gravitino, we will denote this spin connection $\omega(e, \psi)$. The first and second equations are then the standard trivial equations of motion for the graviton and gravitino in 3D.

\subsection{Conformal Supergravity}

Conformal Supergravity can be constructed as the Chern-Simons theory of the 3D $\mathcal{N} = 1$ superconformal algebra, $Osp(1|4)$\cite{VanNieuwenhuizen:1985ff,Uematsu:1986,Foussats:1992}, just as Conformal Gravity has been constructed as the Chern-Simons theory of the conformal group \cite{Horne:1988jf}. As a Chern-Simons theory, Conformal Supergravity has a field corresponding to each transformation generator in the $\mathcal{N} = 1$ superconformal algebra. The translations and Lorentz rotations give rise to $e^a$ and $\omega^a$ respectively, and the supersymmetry transformations correspond to $\psi$. There are also special conformal transformations, conformal supersymmetry transformations and dilatations, so we must introduce new 1-form fields $f^a$, $\phi$ (an anti-commuting Majorana spinor) and $b$ associated to these.

To construct the action, we have followed the working of \cite{Foussats:1992}, and explicitly written out their expression (3.1) in terms of the basic forms

\bea
L_{CSG} &=& \frac{1}{2}\omega_a d\omega^a + \frac{1}{6}\epsilon_{abc}\omega^a\omega^b\omega^c \nonumber \\
&& - 2f_a ( D e^a - b e^a - \bar{\psi}\gamma^a\psi ) \nonumber \\
&& + 4 \bar{\psi}D\phi + \frac{1}{2}b d b - 2b\bar{\psi}\phi - 2 e_a \bar{\phi}\gamma^a \phi \,.
\eea

\begin{table}[h!]
\begin{tabular}{c | c | c | c | c | c | c}
& $P^a$ & $Q$ & $J^a$ & $S$ & $K^a$ & $D$ \\ \hline 
$\delta e^a$ & $(D-b)\xi^a$ & $2 \bar{\epsilon}\gamma^a \psi$ & $-\epsilon^{abc}\lambda_be_c$ & $0$ & $0$ & $\Omega e^a$ \\
$\delta\psi$ & $-\xi^a\gamma_a\phi$ & $(D-\frac{1}{2}b)\epsilon$ & $-\frac{1}{2}\lambda_a\gamma^a\psi$ & $e_a\gamma^a \eta$ & $0$ & $\frac{1}{2} \Omega \psi$ \\
$\delta\omega^a$ & $-2\epsilon^{abc}f_b\xi_c$ & $2\bar{\epsilon}\gamma^a\phi$ & $D\lambda^a$ & $2\bar{\eta}\gamma^a \psi$ & $-2\epsilon^{abc}e_b\chi_c$ & $0$ \\ 
$\delta\phi$ & $0$ & $-f_a\gamma^a \epsilon$ & $-\frac{1}{2}\lambda_a\gamma^a \phi$ & $(D+\frac{1}{2}b)\eta$ & $\chi_a \gamma^a \psi$ & $-\frac{1}{2}\Omega \phi$ \\
$\delta f^a$ & $0$ & $0$ & $-\epsilon^{abc}\lambda_b f_c$ & $-2\bar{\eta}\gamma^a \phi$ & $(D+b)\chi^a$ & $-\Omega f^a$ \\
$\delta b$ & $-2f_a\xi^a$ & $2\bar{\epsilon}\phi$ & $0$ & $-2\bar{\eta}\psi$ & $2e_a\chi^a$ & $d\Omega$ 
\end{tabular}
\caption{Transformations of fields under the superconformal algebra}
\end{table}

To find the equations of motion, first eliminate every variable except $e^a$ and $\psi$ using their equations of motion 

\bea
f&:& \omega_{\mu}\,^a = \omega_{\mu}\,^a(e,\psi,b) \equiv \omega_{\mu}\,^a (e,\psi) + \epsilon_{\mu}\,^{a\nu} b_{\nu} \nonumber \\
\omega&:& R^a - 2 \epsilon^{abc} e_b f_c - 2 \bar{\psi} \gamma^a \phi = 0 \nonumber \\
b&:& db - 2\bar{\psi}\phi + 2e_af^a = 0 \nonumber \\
\phi&:& 4D\psi - 2b\psi + 4e_a\gamma^a\phi = 0 \,.
\eea

Using the Bianchi identity $DDe^a = \epsilon^{abc}R_be_c$, one can see that the $\omega$ and $\phi$ equations imply the $b$ equation. As with Conformal Gravity, this is because $b$ can be gauged away by special conformal transformations, so we can set it to be zero as it must drop out of the final equations of motion. The $\phi$ equation determines $\phi(e,\psi)$ and then the $\omega$ equation defines $f^a(e,\psi)$

\bea
\phi_{\mu} &=& -\frac{1}{2}e^{-1}\gamma_{\nu}\gamma_{\mu}\mathcal{R}^{\nu} \nonumber \\
f_{\mu\nu} &=& \frac{1}{2} S_{\mu\nu}(e) + \rm{fermions} \,,
\eea

\noindent where $e$ is the determinant of the dreibein, $\mathcal{R}^{\mu} = \epsilon^{\mu\nu\rho}D_{\nu}\psi_{\rho}$ and $S_{\mu\nu}(e) \equiv R_{\mu\nu} - \frac{1}{4}R g_{\mu\nu}$ is the 3D Schouten tensor.

The $e^a$ and $\psi$ equations then give the equations of motion for the bosonic and fermionic modes. This action is of course invariant under the superconformal algebra, the action of which is shown in the table above.

\subsection{Topologically Massive Supergravity}

Just as the TMG Lagrangian can be constructed as $-1 \,\times$ the Einstein-Cartan Lagrangian $+\, \frac{1}{\mu}\, \times$ the Conformal Gravity Lagrangian \cite{Hohm:2012}, TMSG can be constructed as $-1\, \times$ the Supergravity Lagrangian $+ \,\frac{1}{\mu} \,\times$ the Conformal Supergravity Lagrangian. This ``Chern-Simons-like'' action, which we will demonstrate is equivalent to the TMSG action given in \cite{Deser:1983}, is

\bea
L_{TMSG} &=& e_aR^a + 2 \bar{\psi}D \psi + \frac{\lambda^2}{6}\epsilon^{abc}e_ae_be_c - \lambda e_a \bar{\psi}\gamma^a \psi + \frac{1}{\mu} \left\{ \frac{1}{2}\omega_a d\omega^a \right. \nonumber \\
&& + \frac{1}{6}\epsilon_{abc}\omega^a\omega^b\omega^c - 2f_a ( D e^a - b e^a - \bar{\psi}\gamma^a\psi ) \nonumber \\
&& \left. + 4 \bar{\psi}D\phi + \frac{1}{2}b d b - 2b\bar{\psi}\phi - 2 e_a \bar{\phi}\gamma^a \phi \right\} \,.
\eea

Now consider the relation between the Euler-Lagrange equations of this theory and those of Chern-Simons Conformal Supergravity above. The TMSG equations are

\bea
f&:& \omega_{\mu}\,^a = \omega_{\mu}\,^a(e,\psi,b) \nonumber \\
\omega&:& R^a - 2 \epsilon^{abc} e_b f_c - 2 \bar{\psi} \gamma^a \phi = -\mu be^a \nonumber \\
b&:& db - 2\bar{\psi}\phi + 2e_af^a = 0 \nonumber \\
\phi&:& 4D\psi - 2b\psi + 4e_a\gamma^a\phi = 0 \,,
\eea

\noindent which differ from the corresponding Conformal Supergravity equations only by the extra term on the right hand side of the $\omega$ equation. Notice that the $\omega$ equation which determines $f^a(e,\psi,\phi,b)$ is linear in $f^a$, so write $f^a = f^a_0 + f^a_1$, where $f^a_0$ is the solution of the Conformal Supergravity equations above. Then the $\omega$ and $b$ equations become

\be
-2\epsilon^{abc}e_bf_{1c} = -\mu be^a \,, \qquad 2e_af_1^a = 0 \,.
\ee

The first equation is equivalent to

\be
f_{1[\mu\nu]} = \frac{1}{2} \mu e^{-1} \epsilon_{\mu\nu\rho}b^{\rho} \,,
\ee

\noindent while the second equation says that the left hand side of this equation vanishes, implying $b = 0$. The first equation then sets $f_1^a = 0$. This implies that $f^a$ is the same as in Conformal Supergravity, and we can determine that $\phi$ is as well. 

Looking ahead to the Hamiltonian formulation, after choosing a time co-ordinate on our space time, the $b_i$ will be dynamical variables, and the equation $b = 0$ indicates the presence of constraints on them. As in \cite{Hohm:2012}, it will make later calculations simpler if we have as few constraints as possible, so we can consider setting $b_{\mu}dx^{\mu} = b_0 dx^0$ in the Lagrangian. Although $b_i = 0$ is a consequence of the equations of motion, this change may in principle alter the theory as it eliminates the two equations of motion of $b_i$, so we must check that the equations of motion are unaltered. Our previous analysis holds until the step where we determine $f_1^a = 0$ and $b = 0$. The single $b_0$ equation becomes

\be
f_{1[ij]} = 0 \,,
\ee

\noindent and the $\omega$ equation becomes

\be
f_{1[ij]} = \frac{1}{2} \mu e^{-1} \epsilon_{ij0}b^0 \,,
\ee

\noindent which implies that $b_0 = 0$ and thus $f_1^a = 0$. This was the only place we used the $b$ equation, so the final equations of motion do not change, and we may set $b_i = 0$ in the Lagrangian without changing the theory.

Now we have expressions for all variables in terms of $e^a$ and $\psi$, the Euler-Lagrange equations for these two variables then give us the equations of motion for Topologically Massive Supergravity, the former being the Topologically Massive Gravity equation with extra fermion terms and the latter an equation for $\psi$. If we substitute the expressions for $\phi$ and $f^a$ into the Lagrangian to make it depend on $e^a$ and $\psi$ only, it becomes

\bea
L_{TMSG} &=& e_aR^a + 2 \bar{\psi}D \psi + \frac{\lambda^2}{6}\epsilon^{abc}e_ae_be_c - \lambda e_a \bar{\psi}\gamma^a \psi \nonumber \\
&& + \frac{1}{\mu} \left\{ \frac{1}{2}\omega_a d\omega^a + \frac{1}{6}\epsilon_{abc}\omega^a\omega^b\omega^c - e^{-1}\bar{\mathcal{R}}^{\mu}\gamma_{\nu}\gamma_{\mu}\mathcal{R}^{\nu} \right\} \,,
\eea

\noindent where $\omega_{\mu}\,^a = \omega_{\mu}\,^a(e, \psi)$ implicitly. This is the same as the usual result\cite{Deser:1983,Deser:1984b,Gibbons:2008} with a cosmological constant\footnote{Our Lagrangians have an overall factor of $\frac{1}{2}$ and our field $\psi$ is scaled by a factor of $\frac{1}{2}$ compared to \cite{Deser:1983}.}.

\section{Hamiltonian Formulation}
\setcounter{equation}{0}

Now that we have first order Lagrangians for our theories, we can easily convert them into Hamiltonian form and then perform an analysis of the Poisson brackets structure of the constraints to determine the number of degrees of freedom of each theory\cite{Dirac:1950pj}. We will introduce the necessary concepts by working through the simpler theories first. The analysis for the purely bosonic theories was performed in \cite{Hohm:2012}.

\subsection{Supergravity}

Consider the theories given by integrating each Lagrangian 3-form we have discussed over a 3-manifold with a Cauchy hypersurface. We will assume the spacetime can be foliated by spacelike surfaces indexed by a time $t$ such that we can decompose our forms as, for example, $e_{\mu}\,^a dx^{\mu} = e_0\,^adt + e_i\,^a d\xi^i$. We transfer our Lagrangian 3-forms into the usual Lagrangians by $L = \mathcal{L}dx^0dx^1dx^2$, getting expressions which are first order in time derivatives for each of our theories.

Our first example, Supergravity, has Lagrangian 

\bea
\mathcal{L}_{SEC} &=& \epsilon^{\mu\nu\rho}\left(-e_{\mu}\partial_{\nu}\omega_{\rho} - \frac{1}{2}\epsilon^{abc}e_{\mu a}\omega_{\nu b}\omega_{\rho c} - 2\bar{\psi}_{\mu}\partial_{\nu}\psi_{\rho} + \omega_{\mu a} \bar{\psi}_{\nu}\gamma^a \psi_{\rho} \right. \nonumber \\
&& \left. -\frac{\lambda^2}{6}\epsilon^{abc}e_{\mu a}e_{\nu b}e_{\rho c} + \lambda e_{\mu a}\bar{\psi}_{\nu}\gamma^a \psi_{\rho} \right) \,.
\eea

To make the necessary calculations easier, rescale the fields as 

\be
e^a \mapsto -e^a\, \qquad \psi \mapsto \frac{1}{2}\psi \,.
\ee

Decompose the spacetime directions into time and spacelike directions 

\be
\mathcal{L}_{SEC} = \epsilon^{ij}e_{ja}\dot{\omega}_i\,^a - \frac{1}{2}\epsilon^{ij}\bar{\psi}_j\dot{\psi}_i + e_{0 a} C_e^a + \omega_{0 a} C_{\omega}^a + \bar{\psi}_0 C_{\psi} \,,
\ee

The fields $e_{0 a}$, $\omega_{0 a}$ and $\psi_0$ can be seen to be non-dynamical, and act as Lagrange multipliers imposing the constraints $C_e^a$, $C_{\omega}^a$ and $C_{\psi}$ respectively, defined as

\bea
C_e^a &=& \epsilon^{ij}(\partial_i\omega_j\,^a + \frac{1}{2}\epsilon^{abc}\omega_{i b}\omega_{j c} + \frac{\lambda^2}{2}\epsilon^{abc}e_{i b}e_{j c} - \frac{\lambda}{4}\bar{\psi}_i \gamma^a \psi_j) \nonumber \\
C_{\omega}^a &=& \epsilon^{ij}(\partial_ie_j\,^a + \epsilon^{abc}e_{i b}\omega_{j c} + \frac{1}{4}\bar{\psi}_i \gamma^a \psi_j) \nonumber \\
C_{\psi} &=& \epsilon^{ij}(-\partial_i\psi_j - \frac{1}{2} \omega_{i a} \gamma^a \psi_j + \frac{\lambda}{2}e_{i a}\gamma^a \psi_j) \,.
\eea

This Lagrangian is now in Hamiltonian form, by which we mean that it is a symplectic term minus a Hamiltonian. The Hamiltonian of a time reparametrisation invariant theory like all those we are working with must vanish, so takes the form of a collection of constraints. The symplectic term tells us the Poisson brackets of the theory

\bea
\left\{\omega_i\,^a(\xi), e_j\,^b(\zeta) \right\} &=& \epsilon_{ij}\eta^{ab}\delta^{(2)}(\xi - \zeta) \nonumber \\ 
\left\{ \psi_i\,^{\alpha}(\xi), \psi_j\,^{\beta}(\zeta) \right\} &=& -\epsilon_{ij}\epsilon^{\alpha \beta} \delta^{(2)}(\xi - \zeta) \,.
\eea

Note that the Poisson brackets of commuting variables are anti-symmetric while the Poisson brackets of anti-commuting variables are symmetric. To work out the number of degrees of freedom the theory has, we must calculated the matrix of Poisson brackets of constraints,

\be
\left\{ C^A_x(\xi), C^B_y(\zeta) \right\}|_{C=0} = P^{AB}_{xy}\delta^{(2)}(\xi - \zeta) \,,
\ee

\noindent where $x, y$ label the fields, $e, \omega, ...$ and $A, B$ a Lorentz index for $e, \omega, f$ a spinor index for $\psi, \phi$ and is absent for $b$. We evaluate this matrix on the constraint surface, where all constraints are set to $0$. The rank of this matrix will then be the number of second class constraints in the theory, the rest being first class. We use the formula ``dimension of physical phase space = dimension of phase space - ($2\,\times$ number of first class constraints) - ($1\,\times$ number of second class constraints)'' to work out the number of physical modes, which is half the dimension of the physical phase space as in \cite{Hohm:2012}.

To compute the Poisson brackets of the constraints, we integrate them against test functions, arbitrary smooth functions with compact support so all surface terms vanish

\be
C(\alpha) = \int d^2 \xi \alpha_a(\xi) C^a(\xi) \,,
\ee

\noindent and find

\bea
\left\{ C_{\omega}(\alpha), C_{\omega}(\beta) \right\} &=& C_{\omega}(\alpha \times \beta) \nonumber \\
\left\{ C_{\omega}(\alpha), C_e(\beta) \right\} &=& C_e(\alpha \times \beta) \nonumber \\
\left\{ C_{\omega}(\alpha), C_{\psi}(\bar{\eta}) \right\} &=& -\frac{1}{2}C_{\psi}(\alpha_a\bar{\eta}\gamma^a) \nonumber \\
\left\{ C_e(\alpha), C_e(\beta) \right\} &=& \lambda^2C_{\omega}(\alpha \times \beta) \nonumber \\
\left\{ C_e(\alpha), C_{\psi}(\bar{\eta}) \right\} &=& \frac{\lambda}{2} C_{\psi}(\alpha_a \bar{\eta} \gamma^a) \nonumber \\
\left\{ C_{\psi}(\bar{\eta}), C_{\psi}(\bar{\xi} \right\} &=& -\frac{1}{2}C_e(\bar{\xi}\gamma^a\eta) + \frac{\lambda}{2}C_{\omega}(\bar{\xi}\gamma^a \eta) \,.
\eea

The algebra of Poisson brackets closes, equivalently all Poisson brackets are zero when evaluated on the constraint surface, so all the constraints of the theory are first class. The dimension of the phase space is 16, as $e_i\,^a$ and $\omega_i\,^a$ have 6 components each and $\psi_i$ has 4, and there are 8 constraints as $e_0^a$ and $\omega_0^a$ have 3 components each and $\psi_0$ has 2. The physical phase space therefore has dimension $16 - (2 \times 8) - (1 \times 0) = 0$; as expected there are no propagating modes.

\subsection{Conformal Supergravity}

Now we do the same thing for Conformal Supergravity, which has Lagrangian

\bea
\mathcal{L}_{CSG} &=& \epsilon^{\mu\nu\rho} \left\{\frac{1}{2}\omega_{\mu a}\partial_{\nu}\omega_{\rho}\,^a + \frac{1}{6}\epsilon_{abc}\omega_{\mu}\,^a\omega_{\nu}\,^b\omega_{\rho}\,^c \right. \nonumber \\
&& - 2f_{\mu a}(\partial_{\nu}e_{\rho}\,^a  + \epsilon^{abc}\omega_{\nu b}e_{\rho c} - b_{\nu}e_{\rho}\,^a - \bar{\psi}_{\nu}\gamma^a \psi_{\rho}) \nonumber \\
&& \left. +4\bar{\psi}_{\mu}\partial_{\nu}\phi_{\rho} - 2\omega_{\mu}\bar{\psi}_{\nu}\gamma^a\phi_{\rho} + \frac{1}{2}b_{\mu}\partial_{\nu}b_{\rho} - 2b_{\mu}\bar{\psi}_{\nu}\phi_{\rho} - 2e_{\mu a}\bar{\phi}_{\nu}\gamma^a\phi_{\rho} \right\} \,.
\eea

It will again be convenient to rescale some of the fields to simplify the calculation; change 

\be
f^a \mapsto -\frac{1}{2} f^a \,, \qquad \psi \mapsto \frac{1}{2}\psi \,, \qquad \phi \mapsto \frac{1}{2}\phi \,.
\ee

The Lagrangian then decomposes into

\bea
\mathcal{L}_{CSG} &=& \frac{1}{2}\epsilon^{ij}\omega_{j a}\dot{\omega}_i\,^a + \epsilon^{ij}f_{j a}\dot{e}_i\,^a + \frac{1}{2}\epsilon^{ij}b_j \dot{b}_i + \epsilon^{ij}\psi_j \dot{\phi}_i \nonumber \\
&& + \omega_{0 a}C_{\omega}^a + f_{0 a}C_f^a + e_{0 a}C_e^a + b_0C_b + \bar{\psi}_0C_{\psi} + \bar{\phi}_0C_{\phi} \,,
\eea

\noindent where the constraints are

\bea
C_{\omega}^a &=& \epsilon^{ij}(\partial_i \omega_j\,^a + \frac{1}{2}\epsilon^{abc}\omega_{i b}\omega_{j c} + \epsilon^{abc}e_{i b}f_{j c} - \frac{1}{2}\bar{\psi}_i \gamma^a \phi_j) \nonumber \\
C_f^a &=& \epsilon^{ij}(\partial_i e_j\,^a + \epsilon^{abc}\omega_{i b}e_{j c} - b_ie_j\,^a - \frac{1}{4}\bar{\psi}_i\gamma^a\psi_j) \nonumber \\
C_e^a &=& \epsilon^{ij}(\partial_i f_j\,^a + \epsilon^{abc}f_{i b}\omega_{j c} + b_i f_j\,^a - \frac{1}{2}\bar{\phi}_i \gamma^a \phi_j) \nonumber \\
C_b &=& \epsilon^{ij}(\partial_i b_j -e_{i a}f_j\,^a - \frac{1}{2}\bar{\psi}_i\phi_j) \nonumber \\
C_{\psi} &=& \epsilon^{ij}(\partial_i \phi_j + \frac{1}{2}f_{i a}\gamma^a \psi_j + \frac{1}{2}\omega_{i a}\gamma^a\phi_j + \frac{1}{2} b_i \phi_j) \nonumber \\
C_{\phi} &=& \epsilon^{ij}(\partial_i \psi_j + e_{i a}\gamma^a\phi_j + \frac{1}{2}\omega_{i a}\gamma^a \psi_j - \frac{1}{2}b_i \psi_j) \,.
\eea

We can compute the Poisson brackets as well, which have been conveniently normalised by the rescaling done earlier

\bea
\left\{ \omega_i\,^a(\xi), \omega_j\,^b(\zeta) \right\} &=& \epsilon_{ij}\eta^{ab} \delta^{(2)}(\xi - \zeta) \nonumber \\
\left\{ e_i\,^a(\xi), f_j\,^b(\zeta) \right\} &=& \epsilon_{ij}\eta^{ab} \delta^{(2)}(\xi - \zeta) \nonumber \\
\left\{ b_i(\xi), b_j(\zeta) \right\} &=& \epsilon_{ij} \delta^{(2)}(\xi - \zeta) \nonumber \\
\left\{ \psi_i\,^{\alpha}(\xi), \phi_j\,^{\beta}(\zeta) \right\} &=& \epsilon_{ij}\epsilon^{\alpha\beta} \delta^{(2)}(\xi - \zeta) \,.
\eea

Again use the test function method to calculate the Poisson brackets of the constraints

\bea
\left\{ C_{\omega}(\alpha), C_{\omega}(\beta) \right\} = C_{\omega}(\alpha \times \beta) \,,&& \qquad
\left\{ C_{\omega}(\alpha), C_f(\beta) \right\} = C_f(\alpha \times \beta) \nonumber \\
\left\{ C_{\omega}(\alpha), C_e(\beta) \right\} = C_e(\alpha \times \beta) \,,&& \qquad
\left\{ C_{\omega}(\alpha), C_b(\sigma) \right\} = 0 \nonumber \\
\left\{ C_{\omega}(\alpha), C_{\psi}(\bar{\eta}) \right\} = -\frac{1}{2}C_{\psi}(\alpha_a\bar{\eta}\gamma^a) \,,&& \qquad
\left\{ C_{\omega}(\alpha), C_{\phi}(\bar{\eta}) \right\} = -\frac{1}{2}C_{\phi}(\alpha_a\bar{\eta}\gamma^a) \nonumber \\
\left\{ C_f(\alpha), C_f(\beta) \right\} = 0 \,,&& \qquad
\left\{ C_f(\alpha), C_e(\beta) \right\} = C_{\omega}(\alpha \times \beta) + C_b (\alpha \cdot \beta) \nonumber \\
\left\{ C_f(\alpha), C_b(\sigma) \right\} = -C_f(\alpha^a \sigma) \,,&& \qquad
\left\{ C_f(\alpha), C_{\psi}(\bar{\eta}) \right\} = -\frac{1}{2}C_{\phi}(\alpha_a\bar{\eta}\gamma^a) \nonumber \\
\left\{ C_f(\alpha), C_{\phi}(\bar{\eta}) \right\} = 0 \,,&& \qquad
\left\{ C_e(\alpha), C_e(\beta) \right\} = 0 \nonumber \\
\left\{ C_e(\alpha), C_b(\sigma) \right\} = C_e(\alpha^a\sigma) \,,&& \qquad
\left\{ C_e(\alpha), C_{\psi}(\bar{\eta}) \right\} = 0 \nonumber \\
\left\{ C_e(\alpha), C_{\phi}(\bar{\eta}) \right\} = -C_{\psi}(\alpha_a \bar{\eta} \gamma^a) \,,&& \qquad
\left\{ C_b(\sigma), C_b(\tau) \right\} = 0 \nonumber \\
\left\{ C_b(\sigma), C_{\psi}(\bar{\eta}) \right\} = -\frac{1}{2}C_{\psi}(\sigma\bar{\eta}) \,,&& \qquad
\left\{ C_b(\sigma), C_{\phi}(\bar{\eta}) \right\} = \frac{1}{2}C_{\phi}(\sigma\bar{\eta}) \nonumber \\
\left\{ C_{\psi}(\bar{\eta}), C_{\psi}(\bar{\xi}) \right\} = \frac{1}{2}C_e(\bar{\xi}\gamma^a\eta) \,,&& \qquad
\left\{ C_{\psi}(\bar{\eta}), C_{\phi}(\bar{\xi}) \right\} = \frac{1}{2}C_{\omega}(\bar{\xi}\gamma^a\eta) - \frac{1}{2}C_b(\bar{\xi}\eta) \nonumber \\
\left\{ C_{\phi}(\bar{\eta}), C_{\phi}(\bar{\xi}) \right\} = C_f(\bar{\xi}\gamma^a\eta) \,.
\eea

The algebra closes again, so all constraints are first class and Conformal Supergravity has $28 - (2\,\times 14) - (1\, \times 0) = 0$ propagating modes, as expected.

\subsection{Topologically Massive Supergravity}

The Topologically Massive Supergravity action, setting $b_i = 0$ as we have earlier checked does not alter the theory, is

\bea
\mathcal{L}_{TMSG} &=& \epsilon^{\mu\nu\rho} \left\{e_{\mu}\partial_{\nu}\omega_{\rho} + \frac{1}{2}\epsilon^{abc}e_{\mu a}\omega_{\nu b}\omega_{\rho c} + 2\bar{\psi}_{\mu}\partial_{\nu}\psi_{\rho} - \omega_{\mu a} \bar{\psi}_{\nu}\gamma^a \psi_{\rho} \right. \nonumber \\
&& \left. + \frac{\lambda^2}{6}\epsilon^{abc}e_{\mu a}e_{\nu b}e_{\rho c} - \lambda e_{\mu a}\bar{\psi}_{\nu}\gamma^a \psi_{\rho} \right\} \nonumber \\
&& + \frac{1}{\mu}\epsilon^{\mu\nu\rho} \left\{\frac{1}{2}\omega_{\mu a}\partial_{\nu}\omega_{\rho}\,^a + \frac{1}{6}\epsilon_{abc}\omega_{\mu}\,^a\omega_{\nu}\,^b\omega_{\rho}\,^c \right. \nonumber \\
&& - 2f_{\mu a}(\partial_{\nu}e_{\rho}\,^a  + \epsilon^{abc}\omega_{\nu b}e_{\rho c} - b_{\nu}e_{\rho}\,^a - \bar{\psi}_{\nu}\gamma^a \psi_{\rho}) \nonumber \\
&& \left. +4\bar{\psi}_{\mu}\partial_{\nu}\phi_{\rho} - 2\omega_{\mu}\bar{\psi}_{\nu}\gamma^a\phi_{\rho} - 2b_{\mu}\bar{\psi}_{\nu}\phi_{\rho} - 2e_{\mu a}\bar{\phi}_{\nu}\gamma^a\phi_{\rho} \right\} \,.
\eea

The Poisson bracket structure is not affected by an overall factor in the Lagrangian so consider the Lagrangian $\mu \mathcal{L}$, the case $\mu = 0$ is Conformal Supergravity which we have already dealt with. Redefine fields

\bea
\tilde{e}^a = \mu e^a \,, \qquad \Omega^a &=& \omega^a + \mu e^a \,, \qquad k^a = \frac{1}{\mu}(-2f^a - \frac{1}{2}\mu^2e^a)\,, \nonumber \\ 
\tilde{\psi} &=& \sqrt{\mu} \psi \,, \qquad \chi = \frac{1}{\mu}(4\phi + 2\mu\psi) \,,
\eea

\noindent and then drop the tildes we have introduced for convenience.

After rewriting everything in terms of these new variables and decomposing the spacetime directions, the Lagrangian becomes

\bea
\mu\mathcal{L}_{TMSG} &=& \frac{1}{2}\epsilon^{ij}\Omega_{j a}\dot{\Omega}_i\,^a + \epsilon^{ij}k_{j a}\dot{e}_i\,^a + \epsilon^{ij}\bar{\chi}_j \psi_j \nonumber \\
&& + \Omega_{0 a}C_{\Omega}^a + k_{0 a}C_k^a + e_{0 a}C_e^a + b_0C_b \bar{\psi}_0C_{\psi} + \bar{\chi}_0C_{\chi} \,,
\eea

\noindent where the constraints are, writing $l = \frac{\lambda}{\mu}$

\bea
C_{\Omega}^a &=& \epsilon^{ij}\left[\partial_i \Omega_j\,^a + \frac{1}{2} \epsilon^{abc}\Omega_{i b}\Omega_{j c} + \epsilon^{abc}e_{i b}k_{j c} - \frac{1}{2}\bar{\psi}_i \gamma^a \chi_j\right] \nonumber \\
C_k^a &=& \epsilon^{ij}\left[\partial_ie_j\,^a + \epsilon^{abc}\Omega_{i b}e_{jc} - \epsilon^{abc}e_{i b}e_{j c} - \bar{\psi}_i\gamma^a\psi_j\right] \nonumber \\
C_e^a &=& \epsilon^{ij}\left[\partial_i k_j\,^a - \frac{1}{2}(1-l^2) \epsilon^{abc} e_{i b}e_{j c} + \epsilon^{abc}k_{i b}\Omega_{j c} - 2 \epsilon^{abc} e_{i b}k_{j c} \right. \nonumber \\
&& \left. - (1+l)\bar{\psi}_i\gamma^a\psi_j + \bar{\psi}_i\gamma^a\chi_j - \frac{1}{8}\bar{\chi}_i\gamma^a\chi_j\right] \nonumber \\
C_b &=& \epsilon^{ij}\left[-e_{i a}k_j\,^a - \frac{1}{2}\bar{\psi}_i\chi_j\right] \nonumber \\
C_{\psi} &=& \epsilon^{ij}\left[\partial_i\chi_j + 2k_{i a}\gamma^a\psi_j + 2(1+l)e_{i a}\gamma^a\psi_j + \frac{1}{2}\Omega_{i a}\gamma^a\chi_j - e_{i a}\gamma^a\chi_j\right] \nonumber \\
C_{\chi} &=& \epsilon^{ij}\left[\partial_i\psi_j + \frac{1}{2}\Omega_{i a}\gamma^a\psi_j - e_{i a}\gamma^a\psi_j + \frac{1}{4}e_{i a}\gamma^a\chi_j\right] \,.
\eea

The complicated field redefinitions earlier were designed to simplify the Poisson brackets as much as possible

\bea
\left\{ \Omega_i\,^a(\xi), \Omega_j\,^b(\zeta) \right\} &=& \epsilon_{ij}\eta^{ab} \delta^{(2)}(\xi - \zeta) \nonumber \\
\left\{ e_i\,^a(\xi), k_j\,^b(\zeta) \right\} &=& \epsilon_{ij}\eta^{ab} \delta^{(2)}(\xi - \zeta) \nonumber \\
\left\{ \psi_i\,^{\alpha}(\xi), \chi_j\,^{\beta}(\zeta) \right\} &=& \epsilon_{ij}\epsilon^{\alpha\beta} \delta^{(2)}(\xi - \zeta) \,.
\eea

We can now work out the Poisson brackets of the constraints. As we are interested in whether the constraints are first or second class, for brevity we shall ignore all multiples of constraints which appear on the right hand side of these equations, equivalently we shall work out the Poisson brackets evaluated on the constraint surface.

\bea
\left\{ C_{\Omega}(\alpha), C_{\Omega}(\beta) \right\} = 0 \,,&& \qquad
\left\{ C_{\Omega}(\alpha), C_k(\beta) \right\} = 0 \,, \nonumber \\
\left\{ C_{\Omega}(\alpha), C_e(\beta) \right\} = 0 \,,&& \qquad
\left\{ C_{\Omega}(\alpha), C_b(\sigma) \right\} = 0 \,, \nonumber \\
\left\{ C_{\Omega}(\alpha), C_{\psi}(\bar{\eta}) \right\} = 0 \,,&& \qquad
\left\{ C_{\Omega}(\alpha), C_{\chi}(\bar{\eta}) \right\} = 0 \,, \nonumber
\eea

\bea
\left\{ C_k(\alpha), C_k(\beta) \right\} &=& -\int d^2\zeta\, \alpha_a\beta_b \epsilon^{ij}e_i\,^ae_j\,^b \,,\nonumber\\
\left\{ C_k(\alpha), C_e(\beta) \right\} &=& \int d^2\zeta\, \alpha_a\beta_b \epsilon^{ij}e_i\,^ak_j\,^b \,,\nonumber \\
\left\{ C_k(\alpha), C_b(\sigma) \right\} &=&  -\int d^2\zeta \, \epsilon^{ij}\left[\partial_i\sigma \alpha_a e_j\,^a - \sigma \alpha_a \epsilon^{abc}e_{i b}e_{j c}\right]  \,,\nonumber \\
\left\{ C_k(\alpha), C_{\psi}(\bar{\eta}) \right\} &=& \int d^2\zeta\, \frac{1}{2}\epsilon^{ij}\alpha_a \bar{\eta}e_i\,^a \chi_j \,, \nonumber \\
\left\{ C_k(\alpha), C_{\chi}(\bar{\eta}) \right\} &=& -\int d^2\zeta\, \frac{1}{2}\epsilon^{ij}\alpha_a \bar{\eta}e_i\,^a \psi_j \,, \nonumber
\eea

\bea
\left\{ C_e(\alpha), C_e(\beta) \right\} &=& -\int d^2\zeta\, \alpha_a\beta_b \epsilon^{ij} \left[k_i\,^ak_j\,^b + (1+l)^2\epsilon^{abc}(\bar{\psi}_i\gamma_c\psi_j \right. \nonumber \\
&& \left. - (1+l) \bar{\psi}_i\gamma_c\chi_j + \frac{1}{4}\bar{\chi}_i\gamma_c\chi_j)\right]   \,, \nonumber \\
\left\{ C_e(\alpha), C_b(\sigma) \right\} &=&  \int d^2\zeta\, \epsilon^{ij}\left[ \partial_i\sigma \alpha_a k_j\,^a + \epsilon^{abc}\sigma\alpha_a\left(\frac{3}{2}(1-l^2)e_{i b}e_{j c} + 2e_{i b}k_{j c}\right) \right. \nonumber \\
&& \left. + 2(1+l) \sigma\alpha_a\bar{\psi}_i \gamma^a \psi_j - \sigma \alpha_a \bar{\psi}_i \gamma^a \chi_j \right] \,, \nonumber \\
\left\{ C_e(\alpha), C_{\psi}(\bar{\eta}) \right\} &=&  \int d^2\zeta\, \frac{1}{2}\epsilon^{ij}\alpha_a \bar{\eta}\left[-k_i\,^a \chi_j - 4(1+l)^2\epsilon^{abc}e_{i b}\gamma_c \psi_j + 2(1+l)\epsilon^{abc}e_{i b}\gamma_c \chi_j\right] \,,  \nonumber \\
\left\{ C_e(\alpha), C_{\chi}(\bar{\eta}) \right\} &=&  \int d^2\zeta\, \frac{1}{2}\epsilon^{ij}\alpha_a \bar{\eta}\left[k_i\,^a \psi_j + 2(1+l)\epsilon^{abc}e_{i b}\gamma_c \psi_j - \epsilon^{abc}e_{i b}\gamma_c \chi_j\right] \,,\nonumber
\eea

\bea
\left\{ C_b(\sigma), C_{\psi}(\bar{\eta}) \right\} &=& \int d^2\zeta\, \epsilon^{ij}\left[ -\frac{1}{2}\partial_i \sigma \bar{\eta}\chi_j + 4(1+l)e_{i a}\sigma \bar{\eta}\gamma^a \psi_j - e_{i a}\sigma\bar{\eta}\gamma^a\chi_j \right] \,, \nonumber \\
\left\{ C_b(\sigma), C_{\chi}(\bar{\eta}) \right\} &=& \int d^2\zeta\, \epsilon^{ij}\left[ \frac{1}{2}\partial_i \sigma \bar{\eta}\psi_j - e_{i a}\sigma \bar{\eta}\gamma^a \psi_j\right] \,, \nonumber \\
\left\{ C_{\psi}(\bar{\eta}), C_{\psi}(\bar{\xi} \right\} &=& \int d^2\zeta\, \epsilon^{ij}\left[-(1+l)^2\epsilon^{abc}e_{i a}e_{j b}\bar{\xi}\gamma_c\eta + \frac{1}{4}\bar{\xi}\chi_i\bar{\eta}\chi_j\right] \,,\nonumber \\
\left\{ C_{\psi}(\bar{\eta}), C_{\chi}(\bar{\xi} \right\} &=& \int d^2\zeta\, \epsilon^{ij}\left[\frac{1}{2}(1+l)\epsilon^{abc}e_{i a}e_{j b}\bar{\xi}\gamma_c\eta - \frac{1}{4}\bar{\xi}\psi_i\bar{\eta}\chi_j\right] \,, \nonumber \\
\left\{ C_{\chi}(\bar{\eta}), C_{\chi}(\bar{\xi} \right\} &=& \int d^2\zeta\, \epsilon^{ij}\left[-\frac{1}{4}\epsilon^{abc}e_{i a}e_{j b}\bar{\xi}\gamma_c\eta + \frac{1}{4}\bar{\xi}\psi_i\bar{\eta}\psi_j\right] \,.
\eea

We want to work out the rank of the $14 \,\times 14$ matrix of Poisson brackets $P^{AB}_{xy}$. Compared to a purely bosonic theory, there is potentially an additional complication as this matrix is a supermatrix. We are trying to find the number of independent second class constraints, which is the number of linearly independent rows or columns of the matrix. A set of vectors $\{v_i\}$ is linearly independent if $\lambda^iv_i = 0 \Rightarrow \lambda^i = 0$. If we allow the $\{v_i\}$ to be Grassmann odd, we can use the same definition as long as we also allow the $\{\lambda^i\}$ to be Grassmann odd, and then everything procedes as in the bosonic case.

To minimise the amount of calculation necessary, first notice that since the three $C^a_{\Omega}$ commute with everything, as in \cite{Hohm:2012} we can separate these conditions out, using them to pick the local frame $e_1\,^a = (0, 1, 0)$, $e_2\,^a = (0, 0, 1)$ and reducing the problem to finding the rank of the $11 \,\times 11$ submatrix, $\hat{P}$ formed by removing the $\Omega$ rows and columns. Next, we compute

\be
\left\{ C_k^0(\xi), C_b(\zeta) \right\} = 2 \delta^{(2)}(\xi - \zeta)\,,
\ee

\noindent and that all other Poisson brackets with $C_k^0$ vanish. The $b$ column is therefore linearly independent of the other columns, and similarly the $b$ row is linearly independent of the other rows. Let the $10 \,\times 10$ submatrix formed by removing the $b$ row and column of $\hat{P}$ be denoted $Q$. Reorder the rows and columns so

\be
\hat{P} = \left(\ba{c c} Q & v \\ w & 0 \ea \right) \, .
\ee 

Now recall that the rank of a matrix, the dimension of its column space $col(M)$ and the dimension of its row space $row(M)$ are all equal. By linear independence of the $b$ row and column

\bea
rank(\hat{P}) &=& dim col \left(\ba{c c} Q & v \\ w & 0 \ea \right) = dim col \left(\ba{c c} Q \\ w \ea \right) + 1 \nonumber \\ 
&=& dim row \left(\ba{c c} Q \\ w \ea \right) + 1 = dim row \left(\ba{c c} Q \ea \right) + 2 = rank(Q) + 2 \,.
\eea

The problem therefore reduces to calculating the rank of $Q$. We have already fixed $e_i\,^a$, and $\Omega_i\,^a$ and $b$ do not appear in $Q$, so the matrix is composed of the $6$ elements of $k_i\,^a$ and the $4$ elements of each of $\psi_i\,^{\alpha}$ and $\chi_i\,^{\alpha}$. After writing the matrix out in terms of these variables, the row space of the matrix can be seen to be spanned by the $k_0^1$, $k_0^2$, $\psi_0^1$ and $\psi_0^2$ rows. More explicitly, first subtract appropriate combinations of the $k_0^1$ and $k_0^2$ rows from all the others in order to remove all elements of $k_i\,^a$ from them, the resulting matrix then has rank 4 by inspection. Therefore, $Q$ has rank $4$, $\hat{P}$ has rank 6 and the original matrix, $P$, also has rank 6. TMSG therefore has a $26 - (2 \, \times 8) - (1 \, \times 6) = 4$ dimensional physical phase space, or equivalently $2$ propagating modes. We know that TMSG is supersymmetric\cite{Deser:1983} so this is a massive spin $\left(2, \frac{3}{2}\right)$ supermultiplet. Note that this is independent of the value of $\lambda$.

\section{Discussion}
\setcounter{equation}{0}

We have constructed a first-order Chern-Simons action for Conformal Supergravity, and then combined this with the Chern-Simons Supergravity action to get a ``Chern-Simons-like'' action for TMSG which is equivalent to the existing formulation. This new action has a number of nice properties, being first-order, metric independent and convenient to work with. We slightly modified this action by setting $b_i = 0$ which we showed does not change the dynamics, a step performed for all dynamically non-trivial ``Chern-Simons-like'' actions so far studied here and in \cite{Hohm:2012}. Using this action we have then shown that TMSG has two propagating modes, which must be a spin $\left(2, \frac{3}{2}\right)$ supermultiplet, while the Chern-Simons theories it is constructed from of course have no local degrees of freedom.

It would be interesting to extend these results to other similar models of massive 3D supergravity\cite{Andringa:2009yc,Bergshoeff:2010mf,Bergshoeff:2010ui}. It was not obvious that a ``Chern-Simons-like'' action for TMSG had to exist, and the existence of ``Chern-Simons-like'' actions for other massive 3D supergravity theories is similarly unclear. One complication is that supersymmetry acts differently on each of the Chern-Simons actions composing TMSG. That the final action is supersymmetric is shown by equivalence to an existing supersymmetric model.

One possible extension would be to $\mathcal{N}=1$ supersymmetric versions of NMG and GMG, a supersymmetrisation of the bosonic ``Chern-Simons-like'' models presented in \cite{Hohm:2012}. Such a theory would involve an extra Lorentz vector 1-form, $k^a$, and one would have to supersymmetrise a term

\be
k_aR^a + \frac{1}{2}\epsilon_{abc}e^ak^bk^c\,.
\ee

One could begin by defining a superpartner $\chi$ to $h^a$, then the first term could be supersymmetrised as the usual Einstein-Hilbert term is by adding $2\bar{\chi}D\chi$. However beyond this the large number of possible extra terms, as well as the existence in the bosonic version of an additional Lorentz scalar 1-form \cite{Hohm:2012}, and the problem of establishing that the final action is supersymmetric make this a complicated task.

Another interesting extension would be to $\mathcal{N}=2$ TMSG. The Supergravity action would then contain a one-form $A$ corresponding to R-Symmetry and perhaps a one-form $C$ corresponding to a central charge, and the Conformal Supergravity action would gain a term $AdA$\cite{Lindstrom:1989,Howe:1996}. The theory would be expected to propagate a spin $\left(2, \frac{3}{2}, \frac{3}{2}, 1\right)$ supermultiplet, the spin-1 mode being formed from $A$ or $C$, but it is difficult to see how any 3-form terms of the type needed for the action to be ``Chern-Simons-like'' could give rise to the required Maxwell action for $A$ or $C$.

\bigskip\bigskip\bigskip

\noindent {\bf Acknowledgements} I would like to thank Paul K. Townsend, at whose suggestion and under whose supervision this work was carried out. I am supported by the STFC.

\vskip .1truecm

\providecommand{\href}[2]{#2}\begingroup\raggedright\endgroup

\end{document}